\documentclass[11pt]{article}
\usepackage[T1]{fontenc}
\usepackage[utf8]{inputenc}
\usepackage[a4paper,margin=1in]{geometry}
\usepackage{newpxtext,newpxmath}
\usepackage{microtype}
\usepackage{amsmath}
\usepackage{booktabs,longtable,tabularx,array}
\usepackage{enumitem}
\usepackage{float}
\usepackage{xcolor}
\usepackage{ragged2e}
\usepackage{tikz}
\usepackage[numbers,sort&compress]{natbib}
\usepackage{xurl}
\usepackage[hidelinks]{hyperref}

\usetikzlibrary{arrows.meta,positioning,fit}
\newcolumntype{Y}{>{\RaggedRight\arraybackslash}X}
\title{Authority--Inference Separation in Agentic Finance:\\First-Line Control, Blockchain Enforcement, and Replayable Assurance}
\author{Hui Gong$^{*}$, Michail Samawi, and Francesca Medda\\
\small UCL Institute of Finance \& Technology, University College London, London WC1E 6BT, UK\\
\small \texttt{h.gong.12@ucl.ac.uk}; \texttt{michail.samawi.25@ucl.ac.uk}; \texttt{f.medda@ucl.ac.uk}\\
\small $^{*}$Corresponding author: \texttt{h.gong.12@ucl.ac.uk}}
\date{August 2026}

\begin{document}
\maketitle

\begin{abstract}
Artificial intelligence agents can select tools, counterparties, and transaction parameters, yet inference should not itself confer authority to execute a financial action. This study develops and evaluates Authority--Inference Separation (AIS), an intent-centered architecture for bounded agentic finance. AIS is scoped to the codifiable subset of first-line preventive control rather than presented as a replacement for institutional governance: risk appetite, mandate approval, agent identity, input governance, and human accountability precede the transaction-time decision, while independent model validation, second-line monitoring, and third-line assurance remain distinct obligations. A financial action intent (FAI) can acquire temporary executable authority only after an independent deterministic control plane validates a persistent agent identifier, accountable owner, mandate and appetite lineage, policy version, state, approvals, and exact economic semantics. A deterministic reason code explains the authorization decision without allowing model rationale to confer authority. Blockchain can then enforce the operational representation of granted authority and record portable settlement evidence. The design distinguishes governance objects that are encodable, attestable, or irreducibly human. Evaluation combines requirements traceability, four-domain instantiation, official BIS and MAS cases, a 48-fixture executable prototype, and a public-ledger observability test. Across 36 synthetic authorization attacks, a direct-agent baseline accepted 36 attack effects, a transparent prompt-policy baseline accepted 20, and AIS accepted none; all configurations accepted 8/8 admissible fixtures. AIS rejected 4/4 token replays and 8/8 recipient or rail substitutions, withheld completion in 4/4 service-delivery failures, and populated all 13 defined evidence fields. These results demonstrate tested mechanisms, not production effectiveness or incident prevalence. The ledger test examines 1,700 recent Base transactions associated with public x402 facilitator addresses, of which 1,193 satisfy a candidate Base-USDC EIP-3009 rule. Public ledgers independently expose settlement, ordering, and selected authorization parameters, but institutional mandate, legal accountability, policy legitimacy, service delivery, and accounting classification require complementary evidence and judgement. A ledger can therefore make granted authority enforceable and auditable across organizational boundaries; it cannot make that authority legitimate.
\end{abstract}


\section{Introduction}

Artificial intelligence (AI) in finance is moving from analysis and recommendation toward workflows that can call tools, select services, prepare orders, initiate payments, and update records. Reasoning-and-acting and tool-use research demonstrates how language models can choose actions and external APIs rather than only generate text \citep{YaoReAct2023,SchickToolformer2023}. In a financial setting, this capability converts a model output into a potential asset transfer, contractual event, limit consumption, or accounting consequence. The institutional question is therefore not simply whether an agent is accurate or authorized. It is how a firm can delegate financially consequential action to adaptive software while preserving authority, operational resilience, and auditability, and who remains accountable when a correctly issued authorization nevertheless produces an adverse outcome \citep{Gong2026A2A,FSB2024,BoEFCA2024}.

The core tension is that inference is probabilistic while execution authority must be determinate at the moment of action. A natural-language instruction may be ambiguous; model behavior may vary with context; and an agent may be exposed to untrusted tool content. Benchmarks already find material safety and prompt-injection failures in tool-using agents \citep{RuanToolEmu2024,DebenedettiAgentDojo2024}. Yet downstream transaction systems must make binary decisions about mandate, eligibility, approval, finality, and record creation. Prompting a model to ``follow policy'' cannot make that policy non-bypassable if the same model or orchestration layer controls the execution path.

AIS is therefore scoped explicitly to the codifiable subset of first-line preventive control. It does not make inherently statistical or ex-post obligations deterministic and does not replace independent model validation, compliance monitoring, internal audit, or governing-body judgement. Instead, it reproduces the transaction-time control architecture through which a human-performed check can migrate to an agent-performed or contract-enforced check while preserving its institutional source, accountable owner, exception path, and evidence. This boundary is a design gain: the first line produces a governed decision population and evidence bundle on which independent monitoring and assurance can operate.

Existing security concepts provide necessary foundations but leave an agentic-finance-specific gap. Role-based access control asks what resources an identity may access; segregation of duties allocates initiation and approval; zero trust emphasizes least-privilege, per-request decisions; reference monitors require complete mediation; and capability systems communicate scoped authority \citep{FerraioloRBAC2001,IIA2020,NISTZTA2020,Anderson1972,SaltzerSchroeder1975,MillerYeeShapiro2003}. None of these concepts by itself defines the machine-generated financial proposal that must be evaluated, bound to temporary authority, reconciled with economic delivery, and replayed for assurance. This article calls that proposal a \textit{financial action intent} (FAI) and makes it the object of control.

Blockchain and tokenisation create an execution problem that is complementary to AIS rather than merely a separate payment opportunity. Smart contracts and tokenised ledgers can enforce transaction conditions, consume nonces with state changes, coordinate a shared settlement state, and provide portable, tamper-evident receipts across organizations. Selected operational representations of institutional control---a mandate registry, quorum approval, time-lock, policy-version hash, or eligible approver set---can also be made programmable. The ledger still cannot establish that the underlying appetite was defensible, an approver understood the decision, consideration was delivered, or accounting treatment is correct. This produces a deliberate division of labour: institutions originate and legitimate authority; AIS decides whether a specific FAI may act; blockchain can enforce the granted scope and make settlement independently verifiable; and governed off-chain evidence supplies mandate, delivery, accounting, and accountability context. Public x402 activity makes both the capability and boundary visible \citep{Ling2026,Wang2026}.

The study addresses three research questions:

\begin{enumerate}[label=\textbf{RQ\arabic*:},leftmargin=*]
\item How should financial authority be separated from probabilistic agent inference?
\item How should first-line control, agent identity, accountability, and evidence handoffs generalize across treasury, trade finance, lending, and investment operations?
\item Which governance objects can blockchain enforce, which can it only witness through attestation, and which authority, delivery, and accounting claims remain irreducibly off-chain or human?
\end{enumerate}

The study makes five contributions. First, it formulates \textit{Authority--Inference Separation} (AIS) as a first-line preventive control that extends reference-monitor, least-privilege, and internal-control principles to a financial action intent while specifying handoffs to the other lines. Second, it defines an institutionally linked FAI, formal authorization invariants, and a governed five-plane architecture spanning appetite and mandate, registered agent identity, inference, deterministic decision, execution, and replayable evidence. Third, it separates model rationale from the deterministic control-decision record and treats only the latter as the auditable explanation of an authorization outcome. Fourth, it classifies programmable governance objects as encodable, attestable, or irreducibly human and shows how blockchain can enforce granted authority without becoming its institutional or legal source. Fifth, it evaluates the artifact through four-domain instantiation, official cases, 48 executable fixtures, and an x402/Base public-ledger test, while preserving explicit boundaries around production effectiveness, model fitness, and adoption.

\section{Related Work and Theoretical Foundations}

\subsection{Agentic Delegation and Tool Use}

AI-enabled digital transformation changes both task execution and the allocation of decision rights. ReAct interleaves language-model reasoning with actions in external environments, while Toolformer demonstrates learned API selection and argument generation \citep{YaoReAct2023,SchickToolformer2023}. Organizational research describes related choices between automation and augmentation and between human and machine decision rights \citep{RaischKrakowski2021,Shrestha2019}. These streams explain why agents can act and how organizations may allocate tasks, but they do not specify a transaction-time mechanism by which an adaptive proposal acquires narrowly bounded financial authority.

An \textit{agentic financial system} is defined here as a software arrangement in which one or more adaptive agents transform goals and observations into financially consequential intents subject to delegated institutional authority. The agent need not be a legal counterparty. Its output is economically consequential when it can influence assets, contractual states, limits, pricing, approvals, accounting entries, or external service procurement.

\subsection{Internal Control, Accounting, and Assurance}

Internal-control frameworks emphasize authorization, segregation of duties, information quality, monitoring, and evidence \citep{COSO2013,IIA2020}. Digital accounting research shows how data-intensive and automated systems expand continuous monitoring and assurance while changing the dependencies of accountants and auditors \citep{Vasarhelyi2015,DaiVasarhelyi2017,Moffitt2018,MollYigitbasioglu2019}. Agentic workflows intensify these issues because the relevant accounting object is not only a final journal entry or payment message. It is the delegated decision chain linking a mandate, FAI, model and tool versions, control evaluation, approval, authorization, execution, service receipt, accounting map, and reconciliation.

This chain divides evidence into pre-action and post-action objects. Pre-action evidence explains why the FAI was admissible: identity, mandate, data provenance, policy version, allowlist or sanctions state, limits, and approvals. Post-action evidence explains what occurred: execution response, settlement identifier, fees, service delivery, accounting classification, reconciliation, and exception status. Preserving only the receipt leaves authority unproven; preserving only the decision leaves execution and economic completion unproven.

The Three Lines Model provides the institutional boundary used here \citep{IIA2020}. AIS is a first-line preventive control owned by management; its evidence bundle is a substrate for second-line monitoring; and deterministic replay is instrumentation that a third line may use without transferring audit responsibility into the system. If complete mediation and evidence capture cover every in-scope FAI, the bundle can make that eligible decision population observable and reduce reliance on sampling. It does not prove that off-system activity is absent, that attestations are true, or that the policy is legitimate; population completeness is itself a control claim that independent functions must test.

\subsection{Reference Monitors, Least Privilege, and Capabilities}

The reference-monitor concept requires every security-relevant access to be mediated by a mechanism that is non-bypassable, tamper resistant, and sufficiently small or structured to be evaluated \citep{Anderson1972}. Complete mediation, fail-safe defaults, separation of privilege, and least privilege remain foundational design principles \citep{SaltzerSchroeder1975}. Zero-trust architecture similarly treats authorization as a per-request decision based on identity, resource, and current context rather than network location \citep{NISTZTA2020}. Capability-based security emphasizes communicating explicit authority rather than relying on broad ambient privilege \citep{MillerYeeShapiro2003}.

AIS adopts these foundations but changes the protected object and the assurance obligation. A conventional reference monitor often decides whether a subject may access an object. AIS decides whether a specific, typed, machine-generated FAI may receive a temporary executable authorization under a current mandate and policy state. The resulting token must be bound to exact action semantics and linked to post-execution evidence. AIS does not claim that mediation, least privilege, or capabilities are new; its novelty is their integration with the FAI and an accounting-grade evidence chain for probabilistic agents.

\subsection{Agent Security, Model Governance, and Operational Resilience}

Tool access turns prompt injection and model error into potential external action. ToolEmu reports high-stakes failure modes in emulated tool environments, and AgentDojo shows that untrusted tool content can hijack realistic agent tasks \citep{RuanToolEmu2024,DebenedettiAgentDojo2024}. These results motivate controls outside the inference context. AI governance also requires validity, reliability, transparency, accountability, privacy, and security \citep{NISTAIRMF2023}. Bounding what a model may execute is nevertheless distinct from establishing that the model is fit for its intended use. Current supervisory guidance treats model identification, approved use, governance, independent validation, monitoring, and material change as a separate risk discipline \citep{PRASS1232026,FedMRM2026}. AIS pins model and data versions and can trigger re-evaluation, but it does not perform independent model validation.

Financial institutions also add expectations for data lineage, risk aggregation, third-party dependency, and recoverability \citep{BCBS239,BCBSResilience2021}. The BoE/FCA survey documents widespread AI use and material third-party dependence \citep{BoEFCA2024}. An agentic control plane should therefore fail closed when material mandate, policy, approval, input quality, or state cannot be verified. A documented degraded or manual route must specify its authorization, evidence, and reconciliation requirements before an outage rather than emerge as an ungoverned bypass.

Identity is likewise more than possession of a signing key. An agent able to consume a limit or move an asset is an actor in the control environment and requires a persistent identifier, registered purpose and permitted domains, lifecycle status, one identity per instance, controlled key custody, and segregation-of-duties checks. Accountability does not transfer to software: the identifier answers which actor acted, while a named natural person and governing body remain responsible for the mandate and its consequences.

\subsection{Programmable Settlement and Selective Verifiability}

Blockchain can provide deterministic state transitions, programmable asset transfer, timestamped receipts, and shared evidence. It can also give AIS decisions operational effect: a blockchain adapter can verify an execution-bound artifact, constrain the permitted state transition, consume a nonce with settlement, and expose a receipt or evidence anchor to parties that do not share an internal log. It also introduces smart-contract, key-management, privacy, concentration, and finality risks. Accounting research notes both the assurance opportunities and the limits of blockchain records \citep{DaiVasarhelyi2017}; economic analysis distinguishes credible commitment from complete institutional governance \citep{CongHe2019}. AIS therefore adopts \textit{selective verifiability} and distinguishes three classes: objects a contract can enforce, claims it can only witness through a signer, and judgements or legal responsibilities that remain human. Operational representations of mandate and policy may be encoded without claiming that a ledger originated or legitimized them.

\subsection{Research Gap: Intent-Centered Authority}

Table~\ref{tab:novelty} locates AIS relative to adjacent mechanisms. The central distinction is that AIS controls neither the identity nor the model in the abstract. It controls whether a particular FAI can acquire temporary executable authority and whether its decision and outcome can be independently reconstructed.

\begin{table}[H]
\caption{AIS compared with adjacent control mechanisms.\label{tab:novelty}}
\small
\begin{tabularx}{\textwidth}{p{0.18\textwidth}p{0.20\textwidth}Y Y}
\toprule
\textbf{Mechanism} & \textbf{Primary object} & \textbf{Typical decision} & \textbf{Gap addressed by AIS}\\
\midrule
IAM / RBAC & Identity, role, resource & May this identity access this resource? & A permitted agent may still propose an impermissible amount, recipient, venue, or economic action.\\
Segregation of duties & Roles in a process & May the same actor initiate and approve? & Role allocation does not bind the exact machine-generated intent or its later execution.\\
AI guardrail & Model input or output & Should the model produce or refuse this content? & A behavioral instruction is not a non-bypassable transaction-time authority decision.\\
Reference monitor / zero trust & Subject--object request & Does this request satisfy access policy now? & Financial actions also require typed economic semantics, settlement, accounting, and delivery evidence.\\
Capability security & Explicit authority object & Does this holder possess scoped authority? & AIS defines how FAI fields, policy versions, expiry, nonce, execution, and assurance are bound.\\
AIS & Financial action intent & May this exact FAI acquire single-use executable authority now? & Integrates adaptive inference with institutional control and replayable financial evidence.\\
\bottomrule
\end{tabularx}
\end{table}

\section{Research Method}

\subsection{Design-Science Sequence and Evaluation Logic}

The study follows design science: problem identification, objective definition, artifact construction, demonstration, and evaluation \citep{Hevner2004,Peffers2007}. Requirements were synthesized from internal control and assurance, the Three Lines Model, AI and model governance, agent security, operational resilience, official programmable-finance projects, and observable x402/Base activity. The artifact comprises the FAI, formal invariants, a governed five-plane reference architecture, evidence and handoff objects, an encodable--attestable--human taxonomy, and a deterministic prototype.

Evaluation uses five complementary forms of evidence. Requirements traceability tests conceptual fit; cross-domain instantiation tests generality; official cases test relevance to institutional programmable-finance workflows; executable fixtures test specified control mechanisms; and the public-ledger module tests the boundary of independently observable evidence. None of these constitutes a production field experiment or a causal estimate of institutional outcomes.

\subsection{Design Requirements}

The synthesis produced eight requirements: (1) inference must not possess standing transaction authority; (2) each acting agent must have a persistent registered identity, declared purpose, lifecycle status, and named accountable human owner; (3) mandates, policies, and limits must link to an approved appetite and change lifecycle, while input provenance and quality are governed before an FAI is formed; (4) every financially consequential execution path must be mediated and authorization must bind exact FAI fields, current policy, expiry, and a single-use nonce; (5) unavailable material state must fail closed or enter a recorded human gate with an eligible and independent approver set; (6) each denial or escalation must emit a deterministic reason code distinct from non-authoritative model rationale; (7) completion must depend on delivery and reconciliation, with breaks aged, escalated, provisioned, and resolved by a named authority; and (8) a logically independent component must be able to replay the decision from pinned inputs and test the completeness of the in-scope evidence population.

The executable prototype tests transaction-time portions of Requirements 1 and 4--8 under fixed fixtures and records the v2 governance fields used by Requirements 2 and 3. It does not test the substantive fitness of an accountable owner, appetite, model, monitoring plan, audit opinion, or attestation. These remain institutional design requirements and handoffs rather than experimentally demonstrated outcomes.

\subsection{Case Selection}

Cases were selected for variation in asset, institution, and workflow while retaining primary documentation. BIS Project Mandala addresses compliance-by-design in cross-border transactions \citep{BISMandala2024}; Project Promissa covers the lifecycle of tokenised promissory notes \citep{BISPromissa2025}; Project Agor\'{a} examines programmable wholesale cross-border settlement \citep{BISAgora2026}; and MAS Project Guardian reports institutional tokenisation use cases across asset management, foreign exchange, and fixed income \citep{MASGuardian2024}. They are experiments, proofs of concept, or programmes rather than evidence of production-scale control effectiveness.

\subsection{Executable Prototype and Fixtures}

The deterministic prototype implements three transparent configurations. \textit{Direct agent} permits an agent to reach the execution adapter without an independent policy decision. \textit{Prompt policy} performs common pre-execution checks in the orchestration layer but has no policy-version pinning, execution-bound token, post-check mutation mediation, or nonce consumption. \textit{AIS} validates a typed FAI in an independent control plane and issues a signed token bound to the canonical intent hash, registered agent, mandate and approval reference, appetite and policy versions, rail, expiry, and nonce; the execution adapter independently verifies that token. The v2 decision record contains a machine-readable reason code, and a delivery break includes age, escalation threshold, provisioning status, and named write-off authority.

The fixture population remains fixed at four financial domains by twelve scenarios, for 48 fixtures and 144 architecture--fixture results. Each domain includes two admissible cases, nine authorization attacks (amount overflow, asset scope violation, recipient substitution, expired mandate, stale policy, missing approval, prompt injection, rail substitution, and replay), and one post-execution service-delivery failure. The primary denominator is 36 authorization attacks. Secondary denominators are eight admissible fixtures, four replay attempts, eight recipient or rail substitutions, and four service-delivery failures. Evidence completeness is the fraction of 13 defined fields present in each record. Adding governance fields changes that descriptive schema denominator but does not change the scenario population or attack denominator. The v2 code, machine-readable fixtures, event records, results, summary, and protocol are preserved separately from v1 in the replication directory.

This evaluation isolates architecture mechanics; it does not call a live language model or estimate an empirical attack probability. The prompt-policy baseline is a specified research implementation rather than a claim about all prompt controls. Exact zero or one rates are expected when a tested mechanism is mandatory or structurally absent.

\subsection{Public-Ledger Observability Method}

Three public sources were preserved on 9 August 2026. The x402 website displayed a 30-day headline snapshot of 75.41 million transactions, USD 24.24 million volume, 94.06 thousand buyers, and 22 thousand sellers \citep{x402Website2026}. A community Dune dashboard supplied cached facilitator-address activity counts, not independently verified x402 settlements. Finally, the Base Blockscout API retrieved up to two latest-activity pages for 18 public, community-listed facilitator addresses; Base lists Blockscout as a supported explorer \citep{BaseExplorers2026}.

The raw sample contains 1,700 address--transaction rows. A candidate was retained only when the transaction completed successfully, called the Base USDC contract, and used EIP-3009 \texttt{transferWithAuthorization} or \texttt{receiveWithAuthorization}. Decoded fields provide payer, recipient, amount, validity bounds, and nonce. The test asks which AIS evidence objects can be verified from those public records, not whether the records prove autonomous-agent adoption.

\section{Authority--Inference Separation}

\subsection{The Financial Action Intent as the Object of Control}

AIS requires the adaptive agent to emit an FAI rather than an opaque tool call. The FAI converts natural-language planning into typed economic semantics while retaining institutional origin, input provenance, and evidence expectations. Table~\ref{tab:fai-main} presents the minimum groups; a production schema would add domain-specific fields and validation rules.

\begin{table}[H]
\caption{Minimum governance, action, and evidence groups in a financial action intent.\label{tab:fai-main}}
\small
\begin{tabularx}{\textwidth}{p{0.19\textwidth}p{0.30\textwidth}Y}
\toprule
\textbf{Field group} & \textbf{Illustrative fields} & \textbf{Authority or assurance purpose}\\
\midrule
Actor identity and mandate & Principal, persistent agent ID, declared purpose, role, mandate ID/version & Names the actor and links the proposal to authenticated delegation without treating software as the authority source.\\
Accountability & Named human owner, approving body, approval reference, next review date & Links the controlled object to the person and body that answer for the mandate.\\
Appetite and policy lineage & Appetite statement ID, tolerance band, breach class, policy version & Records the institutional origin of limits and distinguishes escalation from hard blocking.\\
Action semantics & Domain, action, asset, amount, recipient, venue or rail & Makes the exact economic proposal machine-testable and bindable.\\
Constraints and approvals & Limits, jurisdiction, expiry, eligible human gate, checker independence & Expresses conditions inherited from mandate and policy.\\
Input provenance and model rationale & Model, prompt, tools, trusted/untrusted sources, data versions, alternatives & Preserves diagnostic context as non-authoritative evidence and makes input quality visible on replay.\\
Control decision and integrity & State snapshot, passed/failed rules, reason code, delivery, accounting map, serialization, hash, nonce & Provides the authoritative authorization record and connects control, execution, reconciliation, and replay.\\
\bottomrule
\end{tabularx}
\end{table}

The identity and accountability rows are deliberately separate. A key or address can show which credential acted, but a registered agent identity supplies purpose, permitted domains, lifecycle, and segregation-of-duties context. Neither becomes a legal person. The agent identifier answers which actor acted; the accountable owner and approving body answer who is responsible for the delegated authority and its consequences.

The same separation resolves explainability without allowing explanation to grant authority. Model rationale is descriptive, probabilistic, and useful for diagnosis and challenge. The control-decision record is deterministic, reproducible from pinned inputs, and authoritative for the authorization outcome. It should emit a reason code for every denial and escalation. This \textit{authorization explanation of record} does not necessarily satisfy every explanation obligation for the underlying economic decision, particularly where a model, human judgement, or protected legal right determines the substantive outcome.

\subsection{Formal Decision and Architecture Invariants}

Let an agent generate intent $I$ from context $X$ under model state $M$. Let $G$ be the authenticated governance linkage comprising registered agent identity, accountable owner, mandate approval, and appetite lineage; $A$ an active mandate; $P_v$ a versioned policy set; $Q$ a governed input-quality record; $S_t$ the relevant financial and operational state at time $t$; and $H$ any required human approval. The independent control function returns either a scoped authorization token $\tau$ or denial $\bot$ with deterministic reason code $r$:

\begin{equation}
C(I,G,A,P_v,Q,S_t,H) \in \{(\tau,r),(\bot,r)\}.
\end{equation}

Execution is accepted only when the token is authentic, bound to the canonical intent hash, unexpired, and unused:

\begin{equation}
\begin{aligned}
\mathrm{Execute}(I,t)=1 \Longleftrightarrow {} & \mathrm{Verify}(\tau)=1
\land \mathrm{Bind}(\tau,h(I))=1 \\
& \land t<t_{\mathrm{exp}} \land \mathrm{Unused}(n_{\tau})=1.
\end{aligned}
\end{equation}

After an accepted execution, nonce consumption is atomic with the adapter decision:

\begin{equation}
\mathrm{Execute}(I,t)=1 \Rightarrow \mathrm{Used}(n_{\tau}) \leftarrow 1.
\end{equation}

These rules yield seven architecture invariants: (I1) inference does not imply authority; (I2) the acting identity, accountable owner, and institutional authority lineage are explicit; (I3) authorization is bound to the exact FAI; (I4) every state-changing path performs complete mediation; (I5) authority is scoped, short-lived, and single-use; (I6) every control outcome has a deterministic reason record; and (I7) decision, execution, delivery, and reconciliation evidence is independently replayable. They are design obligations, not mathematical proof that an implementation, attestation, or policy is correct.

\begin{table}[H]
\caption{AIS invariants, mechanisms, and evaluation status.\label{tab:invariants}}
\footnotesize
\begin{tabularx}{\textwidth}{p{0.09\textwidth}p{0.25\textwidth}Yp{0.24\textwidth}}
\toprule
\textbf{Invariant} & \textbf{Requirement} & \textbf{Mechanism} & \textbf{Evaluation status}\\
\midrule
I1 & $\mathrm{Inference}(I)\not\Rightarrow\mathrm{Authority}(I)$ & Independent deterministic decision & Tested against direct and prompt-context execution.\\
I2 & Named actor and authority lineage & Registry, accountable owner, mandate approval, appetite and policy references & Recorded and bound in v2; substantive legitimacy not tested.\\
I3 & Exact FAI binding & Canonical serialization and signed intent hash & Recipient, amount, asset, and rail effects tested.\\
I4 & Non-bypassability & Verification at every execution adapter & Direct and alternate post-check paths tested.\\
I5 & Ephemeral single use & Expiry and atomic nonce consumption & Four replay attempts tested.\\
I6 & Deterministic authorization explanation & Passed/failed rules and reason code & Recorded in all AIS control decisions; legal sufficiency not tested.\\
I7 & Replayable and reconcilable evidence & Pinned governance, input, state, decision, receipt, delivery, and break record & Schema and replay tested; truth and audit opinion not tested.\\
\bottomrule
\end{tabularx}
\end{table}

\subsection{Five-Plane Reference Architecture}

\begin{figure}[H]
\centering
\begin{tikzpicture}[
  node distance=3.2mm,
  box/.style={draw=black!65, rounded corners=2pt, minimum width=0.91\textwidth,
              text width=0.86\textwidth, minimum height=8mm, align=center, font=\footnotesize},
  meta/.style={draw=black!70, dashed, rounded corners=2pt, minimum width=0.91\textwidth,
              text width=0.86\textwidth, minimum height=8mm, align=center, font=\footnotesize, fill=gray!9},
  flow/.style={-{Latex[length=2.2mm]}, line width=0.7pt, draw=black!70}
]
\node[meta] (meta) {\textbf{Institutional Governance and Meta-Control}\\Risk appetite; mandate approval; accountable owner; dual-control keys; verifier upgrade and change control; global revocation; continuity};
\node[box, fill=blue!8, below=of meta] (mandate) {\textbf{1. Mandate and Identity Plane}\\Persistent agent identity, purpose, approval reference, delegated scope, limits, review, expiry and revocation};
\node[box, fill=cyan!8, below=of mandate] (infer) {\textbf{2. Adaptive Inference Plane}\\Observation, planning, retrieval, tool selection, recommendation, typed FAI};
\node[box, fill=orange!13, below=of infer] (control) {\textbf{3. Deterministic Authority and Control Plane}\\Input-quality, appetite and policy checks; segregation of duties; reason code; single-use authorization};
\node[box, fill=green!10, below=of control] (execute) {\textbf{4. Execution and Settlement Plane}\\Bank and venue adapters; blockchain adapter and smart-contract verifier; tokenised-asset lifecycle};
\node[box, fill=purple!9, below=of execute] (evidence) {\textbf{5. Evidence and Handoff Plane}\\Decision and receipt; anchor; delivery; accounting; break ageing; second-line monitoring input; third-line replay};
\draw[flow] (meta) -- node[right,font=\scriptsize]{legitimate and govern} (mandate);
\draw[flow] (mandate) -- node[right,font=\scriptsize]{authenticated scope} (infer);
\draw[flow] (infer) -- node[right,font=\scriptsize]{intent, never authority} (control);
\draw[flow] (control) -- node[right,font=\scriptsize]{bound token or denial} (execute);
\draw[flow] (execute) -- node[right,font=\scriptsize]{receipt and state change} (evidence);
\end{tikzpicture}
\caption{Governed reference architecture and division of labour. AIS controls the codifiable first-line path; operational representations of authority may be contract-enforced, while legitimacy, legal accountability, model validation, monitoring, and audit remain institutionally governed.\label{fig:architecture}}
\end{figure}
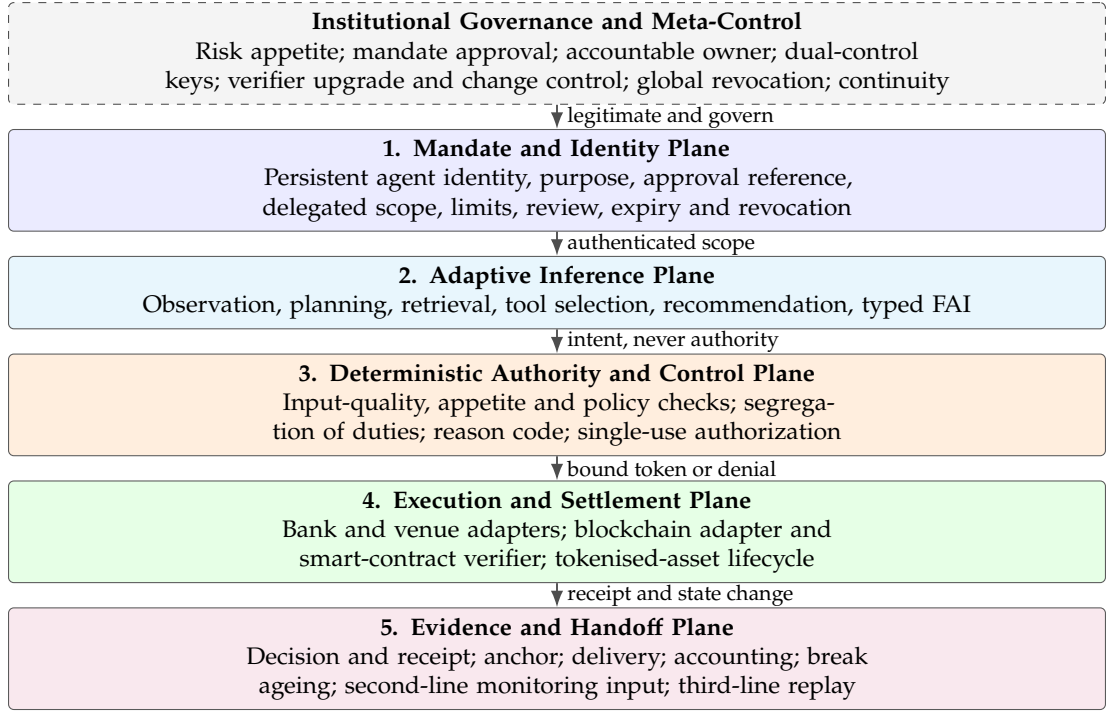

Figure~\ref{fig:architecture} separates adaptive reasoning from any component capable of changing financial state and makes explicit the authorities that govern the architecture itself. Typed artifacts cross each execution boundary; unstructured explanations may accompany an FAI but cannot substitute for it.

The authority plane contains the registered agent and mandate state, policy decision point, approval gate, reason-code service, and token service. Every execution adapter validates the token rather than trusting the agent's self-reported compliance. On a blockchain path, an adapter submits the AIS authorization to a smart-contract verifier that checks the canonical FAI hash, registered agent, mandate and policy references, scope, expiry, recipient, asset, rail, and nonce before the permitted state transition. Nonce consumption should be atomic with settlement, and the resulting event or receipt can carry an evidence hash linking the transaction to the governed off-chain bundle. Tokenised-asset lifecycle events can preserve this linkage through issuance, transfer, servicing, redemption, and accounting.

The meta-control layer is not a sixth execution plane. It governs who may change the controls: custody of signing and upgrade keys, contract and policy change, global mandate revocation, propagation latency, and treatment of in-flight tokens. A material increase can require a higher quorum and time-lock, but those mechanisms prove that configured keys approved a change, not that the change was prudent. Revocation and upgrade authority therefore require the same segregation, accountability, and evidence as transaction authority.

Where the deterministic control plane is unavailable, the architecture fails closed. Operational resilience requires a pre-defined manual or degraded path with its own authorization level, eligible approvers, evidence requirements, and reconciliation back into the AIS chain. After execution, a retained break is not treated as a completed control merely because it was detected: its age, escalation threshold, provisioning status, and named write-off authority remain part of first-line operation. The evidence plane supplies second-line and third-line users without assigning their independent responsibilities to AIS.

\section{Cross-Domain Validation}

\subsection{Financial-Domain Instantiation}

AIS separates authority logic from any single rail so that blockchain can be integrated without treating control of a key or ledger account as institutional mandate. In cross-organizational or tokenised workflows, the blockchain adapter gives distinctive effect to the same FAI by verifying scope, constraining state transitions, settling assets, and exporting portable receipts. Table~\ref{tab:domains} shows how the FAI, authority, human-governance, and evidence objects acquire domain-specific semantics.

\begin{table}[H]
\caption{Cross-domain instantiation of AIS.\label{tab:domains}}
\footnotesize
\begin{tabularx}{\textwidth}{p{0.11\textwidth}p{0.20\textwidth}p{0.24\textwidth}p{0.19\textwidth}Y}
\toprule
\textbf{Domain} & \textbf{Adaptive proposal} & \textbf{Deterministic authority checks} & \textbf{Governance and human gate} & \textbf{Execution and accounting evidence}\\
\midrule
Treasury & Select bank, currency, timing, hedge, or liquidity source; create payment or FX FAI. & Entity mandate, appetite-linked cash limit, approved bank, beneficiary, sanctions state, FX tolerance. & Treasury committee; independent dual approval; periodic entitlement recertification. & Receipt, rate and fee, finality, cash and FX entries, bank reconciliation and aged breaks.\\
Trade finance & Extract terms, identify discrepancies, assemble evidence, propose conditional release. & Instrument version, identities, document completeness, jurisdiction rules, discrepancy approval. & Trade authority matrix; exception forum; independent document reviewer. & Document hashes, release state, liability change, fees, exception and reviewer record.\\
Lending & Collect evidence, propose terms or servicing action, prepare disbursement FAI. & Product eligibility, approved model/version, protected-attribute controls, concentration and delegation limits. & Credit committee; model validation handoff; adverse decisions routed to human review where applicable. & Deterministic authorization reason, decision notice, contract version, disbursement, credit-loss inputs and override history.\\
Investment operations & Interpret mandate, propose order and route, rebalance, or procure data/model service. & Client restrictions, pre-trade compliance, exposure, liquidity, approved venue or vendor. & Investment committee; exception forum; best-execution and conduct monitoring remain ex-post. & Order and fill, timestamps, costs, holdings, cash, allocation, service receipt and reconciliation.\\
\bottomrule
\end{tabularx}
\end{table}

\subsection{Comparative Case Mapping}

The official cases support the relevance of programmable compliance, tokenised obligations, and shared settlement, while showing why institutional and legal controls remain outside the ledger. Table~\ref{tab:cases} maps each case to AIS without treating it as adoption or effectiveness proof.

\begin{table}[H]
\caption{Comparative case evidence and bounded interpretation.\label{tab:cases}}
\small
\begin{tabularx}{\textwidth}{p{0.19\textwidth}YYp{0.23\textwidth}}
\toprule
\textbf{Case} & \textbf{AIS relevance} & \textbf{Primary control object} & \textbf{Interpretive boundary}\\
\midrule
BIS Mandala & Policy requirements can accompany cross-border workflows. & Machine-readable compliance evidence linked to institutional rules. & Experiment; institutions interpret rules and retain responsibility.\\
BIS Promissa & Tokenisation can connect issuance, transfer, payment, and lifecycle records. & Instrument state, authorized holder, event, and accounting evidence. & Proof of concept; legal and operational integration remains.\\
BIS Project Agor\'{a} & Shared programmable infrastructure can coordinate wholesale cross-border payment. & Identity, conditions, atomic settlement, finality, and liquidity. & Controlled testing, not production effectiveness or universal architecture.\\
MAS Project Guardian & Multiple asset classes require interoperable institutional controls. & Participants, asset rules, venue, custody, and settlement evidence. & Use-case descriptions do not provide causal outcome estimates.\\
x402 on Base & Machine-native HTTP payment creates independently inspectable authorization and settlement evidence. & EIP-3009 fields, validity and nonce, facilitator submission, transaction ordering, and receipt. & Strong execution evidence; institutional mandate, independent agency, and delivery still require off-chain proof.\\
\bottomrule
\end{tabularx}
\end{table}

\section{Blockchain Execution and Public-Ledger Evidence Test}

\subsection{Sample Construction and Descriptive Profile}

Of 1,700 deduplicated transactions in the Blockscout sample, 1,507 completed successfully and 193 failed or returned an error status. The EIP-3009 candidate rule retained 1,193 transactions. Decoding identified 173 payer addresses, 25 recipient addresses, and 177 payer--recipient pairs. Candidate value totaled USD 470.866; the median was USD 0.10, the transaction-value Gini coefficient was 0.832, the maximum was USD 25.00, and the most frequent recipient accounted for 60.85\% of candidate count. These values describe a non-random latest-activity sample and are not population estimates.

\begin{table}[H]
\caption{Descriptive profile of the Base EIP-3009 candidate sample.\label{tab:x402sample}}
\small
\begin{tabularx}{\textwidth}{p{0.27\textwidth}p{0.17\textwidth}Y}
\toprule
\textbf{Metric} & \textbf{Value} & \textbf{Interpretation}\\
\midrule
Raw and deduplicated rows & 1,700 & Up to two recent pages for 18 community-listed addresses; not a time-balanced census.\\
Successful rows & 1,507 & Blockscout reports success; unrelated facilitator-address activity may remain.\\
EIP-3009 candidates & 1,193 & Successful Base-USDC authorization calls; narrower than address activity.\\
Unique payers / recipients & 173 / 25 & Address endpoints are not verified economic actors.\\
Total / median value & USD 470.866 / USD 0.10 & Sample value only; not directly comparable with ecosystem headline volume.\\
Value Gini / top recipient share & 0.832 / 60.85\% & Sample concentration; the design prevents population inference.\\
\bottomrule
\end{tabularx}
\end{table}

\subsection{Dual Finding: On-Chain Assurance and Off-Chain Evidence Gaps}

The ledger test asks both what blockchain contributes and where complementary institutional evidence is required. An independent observer can verify successful inclusion, ordering, contract and method, payer, recipient, amount, validity bounds, nonce, and receipt without relying on the facilitator's internal database. These fields are machine-verifiable post-action assurance inputs and, in an AIS implementation, can demonstrate whether execution matched the authorization placed before the adapter. Table~\ref{tab:observability} is therefore read as an evidence specification rather than merely a limitation: every object unavailable on-chain creates a mandatory off-chain evidence obligation whose absence should be detectable.

\begin{table}[H]
\caption{Public-ledger observability across the AIS evidence chain.\label{tab:observability}}
\small
\begin{tabularx}{\textwidth}{p{0.22\textwidth}p{0.15\textwidth}p{0.17\textwidth}Y}
\toprule
\textbf{AIS evidence object} & \textbf{Public chain} & \textbf{x402/Base sample} & \textbf{Off-chain evidence required}\\
\midrule
Institutional mandate & Representation only & No & Approved source, delegation scope, validity, revocation, accountable owner, and legal responsibility.\\
Agent identity and control & Representation only & No & Persistent registry identity, purpose, lifecycle, key custody, operator, and signed delegation.\\
Financial action intent & Usually no & No & Resource, economic purpose, alternatives, quoted price, and canonical FAI.\\
Authorization fields & Partial & Yes & EIP-3009 payer, recipient, amount, validity, and nonce are visible; institutional approval is not.\\
Policy decision & No & No & Policy version, state snapshot, passed and failed rules, and human approval.\\
Settlement & Yes & Yes & Transaction status and receipt are visible, subject to chain-specific finality.\\
Service delivery & No & No & Independent resource response or contractual performance evidence.\\
Accounting classification & No & No & Rights and obligations, valuation, journal map, reconciliation, and presentation.\\
\bottomrule
\end{tabularx}
\end{table}

The positive finding is therefore not merely that a public trace exists. Selected authorization semantics and the resulting settlement state can be checked independently, transported across organizational boundaries, and linked by hash to an AIS decision bundle. A registry or approval representation can also appear on-chain, but its presence demonstrates configured state, not the legitimacy of its institutional source. The boundary finding is that the evidential hierarchy still runs from facilitator-address activity to candidate authorization call, verified protocol settlement, authenticated agent initiation, independently controlled counterparties, and delivered economic service. Each level requires additional evidence. This interpretation is consistent with the population-scale authenticity analysis of Ling et al. \citep{Ling2026} and the facilitator-security analysis of Wang et al. \citep{Wang2026}. A chain receipt is a powerful post-action assurance object, but it complements rather than replaces pre-action mandate, policy, accountability, and delivery evidence.

\section{Executable Prototype Evaluation}

\subsection{Configurations, Metrics, and Validity Boundary}

The prototype tests whether architectural separation makes specified attack effects non-executable. A result is an \textit{unauthorized execution} when the manipulated or replayed effect is accepted in one of the 36 authorization-attack fixtures. For replay fixtures, a legitimate first use seeds the test and only the second use is counted as the attack. A \textit{substitution success} is an accepted recipient or rail mutation after the initial check. An \textit{unsafe completion} occurs when the workflow is marked complete despite missing service delivery. Control-decision replay requires the pinned governance, input, mandate, policy, state, reason code, decision, and outcome needed for independent reconstruction.

Evidence completeness uses 13 equally weighted fields: fixture ID, architecture, canonical intent, agent-registry snapshot, mandate snapshot, appetite linkage, policy version, input-quality record, control decision, authorization artifact or explicit denial, execution outcome or explicit non-execution, reconciliation state, and replay result. This is a schema-completeness measure, not evidence truth, population completeness, policy legitimacy, or institutional audit quality.

\subsection{Results}

Table~\ref{tab:prototype-results} reports exact counts and denominators. All configurations accepted all eight admissible fixtures, so the AIS test did not introduce a false block in this fixed population. Direct access accepted every authorization attack effect. The prompt-policy baseline blocked amount, asset, mandate-expiry, and missing-approval violations, but it accepted stale-policy, injection-override, post-check substitution, and replay effects. AIS rejected all tested unauthorized effects because the adapter verified exact intent binding, active policy, and nonce state outside the inference path.

\begin{table}[H]
\caption{Deterministic prototype results across 48 fixtures per configuration.\label{tab:prototype-results}}
\footnotesize
\begin{tabularx}{\textwidth}{p{0.17\textwidth}p{0.15\textwidth}p{0.11\textwidth}p{0.11\textwidth}p{0.14\textwidth}Y}
\toprule
\textbf{Configuration} & \textbf{Unauthorized execution} & \textbf{Valid accepted} & \textbf{Replay} & \textbf{Substitution} & \textbf{Evidence completeness / replay}\\
\midrule
Direct agent & 36/36 (100\%) & 8/8 & 4/4 & 8/8 & 38.5\% / 0/48\\
Prompt policy & 20/36 (55.6\%) & 8/8 & 4/4 & 8/8 & 53.8\% / 0/48\\
AIS & 0/36 (0\%) & 8/8 & 0/4 & 0/8 & 100\% / 48/48\\
\bottomrule
\end{tabularx}
\end{table}

In the four service-delivery failures, direct-agent and prompt-policy workflows marked 4/4 as complete after execution, whereas AIS marked 0/4 complete and retained a reconciliation break with age, escalation threshold, provisioning status, and named write-off authority. The outcome demonstrates the distinction between authorized execution and evidenced economic completion. It also shows that a blockchain receipt can provide strong settlement assurance while remaining insufficient, by itself, to close the delivery and accounting portions of the AIS evidence chain. Because these are one-period fixtures, the recorded break fields do not estimate operational ageing or escalation rates.

\subsection{Interpretation}

The comparison demonstrates a structural proposition: controls implemented only before or inside an adaptive orchestration context can be separated in time from the actual state-changing call. An execution-bound authorization artifact closes that gap for the fields and attacks represented in the harness. The experiment does not show that every prompt control will have a 55.6\% failure rate, nor that every AIS implementation will achieve zero failures. The rates are conditional on transparent baseline definitions and fixed fixtures. The scientific contribution is the executable mapping from invariants to observable outcomes and machine-readable records, not an estimate of real-world security incidence.

\section{Discussion}

\subsection{Theoretical Contribution}

AIS extends the reference-monitor principle from general resource access to machine-generated financially consequential intent. Its object is the FAI; its authorization is scoped to exact economic semantics; and its first-line evidence obligation spans actor identity, appetite and mandate lineage, policy, input quality, execution, delivery, accounting, and replay. This distinguishes AIS from rebranding conventional access control while avoiding a claim to replace the full governance architecture. IAM and RBAC remain necessary for authenticating principals and services, segregation of duties remains necessary for approval design, second-line specialists retain monitoring and challenge, and internal audit retains independent assurance. AIS specifies the preventive control and evidence handoff through which these roles operate when probabilistic inference proposes a financial action.

The prototype provides mechanism evidence for the transaction-time portion of this claim. The result is not that deterministic code is universally safer than AI, that the registered owner is competent, or that the policy is legitimate. It is that a component cannot reliably police authority it can also bypass. The non-bypassable adapter, exact FAI binding, and reproducible reason record remain meaningful whether inference is supplied by one model, multiple agents, rules, or a human-assisted system.

\subsection{Strategy and Digital-Transformation Implications}

AIS reframes agentic transformation as a governed migration from human-performed to agent-performed controls rather than a model-procurement decision. Firms can change inference providers without silently changing transaction authority because registered actors, mandates, appetite linkage, and policy decisions remain in a separate institutional control plane. The same modularity that supports experimentation also makes the inference provider a critical third party: concentration, substitutability, exit, and degraded operation must be assessed rather than assumed away. Ownership remains explicit: governing bodies set appetite and accountability, business functions own mandates and first-line operation, risk and compliance define or challenge admissibility, technology enforces mediation, finance records consequences, and internal audit independently evaluates the chain.

The architecture also clarifies automation versus augmentation. Low-risk, high-frequency FAIs may pass deterministic controls without human approval; material, novel, or ambiguous FAIs enter a documented exception gate with an eligible approver set independent of the mandate owner. Human participation is therefore triggered by risk and evidenced rather than inserted into every action or removed indiscriminately. Accountability never transfers to the agent: the persistent identifier establishes traceability, while the named owner and approving body remain answerable for delegation, review, and correction.

\subsection{Accounting and Auditing Implications}

The FAI and its evidence bundle constitute a candidate accounting and assurance object for agentic systems. Traditional logs often preserve user, endpoint, and timestamp but omit registered agent purpose, accountable owner, appetite lineage, model and data provenance, alternatives, mandate approval, failed rules, delivered service, and accounting mapping. Standardizing these fields can support continuous controls monitoring, transaction-level cost attribution, deterministic authorization explanations, independent replay, and reconciliation between operational evidence and journal entries.

For the population covered by complete mediation, the bundle may allow second-line or audit users to test every recorded FAI rather than draw a conventional sample. That claim is conditional: reviewers must establish the population boundary, reconcile execution endpoints to evidence records, and detect off-system or manual activity. Replay proves that a recorded decision conformed to pinned policy; it does not establish that the policy was defensible. Independent assurance must therefore address both evidence completeness and policy legitimacy.

Auditors should treat a blockchain receipt as a powerful but incomplete assurance object. It can independently evidence the executed contract call, selected authorization semantics, transaction ordering, settlement state, and any anchored hash, reducing dependence on a single organization's mutable operational log. An on-chain transfer does not by itself establish institutional rights and obligations, approval, valuation, presentation, or delivered consideration. Selective anchoring can make later alteration detectable without disclosing confidential source data. To reduce tension with applicable retention or erasure obligations, the design anchors hashes rather than personal or commercially sensitive source data and assigns both the off-chain bundle and anchor a stated retention policy.

\subsection{The Appropriate Role of Blockchain}

Blockchain performs four distinctive functions within AIS. First, as an \textit{enforcement substrate}, a smart-contract verifier can accept only an authorization bound to the permitted FAI fields and couple nonce consumption to the state transition. Second, as a \textit{settlement-assurance substrate}, it exposes ordering, contract execution, asset movement, and finality evidence to independent parties. Third, as a \textit{cross-organizational evidence substrate}, it can carry receipts or anchors without requiring one participant's database to become the common source of truth. Fourth, as a \textit{programmable-governance substrate}, it can enforce selected operational representations of institutional control: registry state, quorum approval, authority lookups, higher-quorum appetite parameters, time-locks on material mandate increases, eligible approver sets, policy-version hashes, and conditional release.

These functions do not apply equally to every governance object. Table~\ref{tab:eah} separates what a contract can enforce, what it can only witness through attestation, and what remains irreducibly human.

\begin{table}[H]
\caption{Encodable, attestable, and human governance objects in programmable settlement.\label{tab:eah}}
\footnotesize
\begin{tabularx}{\textwidth}{p{0.12\textwidth}p{0.20\textwidth}p{0.38\textwidth}Y}
\toprule
\textbf{Class} & \textbf{What the ledger can do} & \textbf{Illustrative objects} & \textbf{Boundary}\\
\midrule
E---Encodable & Enforce deterministic state or transition after institutional authorization. & Mandate registry; m-of-n approval; signed authority lookup; higher-quorum parameter; time-lock; agent ID bound to purpose; one key per instance; disjoint checker keys; pinned policy hash; conditional release; evidence anchor. & Proves configured state and signatures, not that the configuration or approval was prudent.\\
A---Attestable & Verify signer, integrity, time, and linkage of a claim whose truth is external to the chain. & Service delivery; document truth; owner acceptance; input-quality threshold; counterparty independence; off-chain state at the decision time. & Establishes that an attestation was made, not that the attested fact was true.\\
H---Human & Record or route a judgement without encoding its substantive correctness or legal responsibility. & Whether appetite or a limit is defensible; whether an approver understood; competence; legal accountability; custody of the upgrade key; correction of a wrong but valid instruction. & Judgement, legal personhood, and the human root of delegated authority cannot be eliminated by recursion into more contracts.\\
\bottomrule
\end{tabularx}
\end{table}

A ledger can therefore make granted authority enforceable; it cannot make that authority legitimate. These functions are most valuable when multiple organizations require a common execution state, tokenised assets require programmable lifecycle controls, or machine-native services need portable receipts. They are less valuable when one institution controls all relevant systems, confidentiality dominates, legal finality is external to the chain, or the decisive facts cannot be encoded reliably. Programmability also raises the governance burden: irreversible settlement narrows correction options, machine speed can propagate a valid but wrong instruction before review, and an upgrade key may concentrate more effective authority than a conventional delegation.

The x402 test makes this division of labour concrete. Public settlement activity exposes authorization and settlement fields that independent parties can inspect, demonstrating blockchain's contribution to portable assurance. The same activity can generate misleading headline metrics because it does not identify institutional mandate or delivered demand. Independent studies report extreme concentration, internal activity, and facilitator vulnerabilities \citep{Ling2026,Wang2026}. Strategy should therefore value verifiable execution while avoiding the equation of transaction count with trusted adoption. Decision-relevant indicators include independently controlled payer--recipient relationships, verified service delivery, repeat use under explicit mandates, exception and escalation rates, aged reconciliation breaks, deterministic reason-code distributions, and evidence completeness.

\subsection{Limitations and Research Agenda}

First, the executable prototype is a deterministic, author-implemented architecture-mechanism test. It contains no live model, production key, concurrency, economic optimization, compromised control plane, key theft, collusion, or implementation vulnerability. Its baseline definitions shape the exact rates. The v2 governance fields are recorded and bound, but the harness does not separately attack agent provisioning, appetite approval, key custody, upgrade authority, revocation propagation, or the manual route. Broader evaluation should add multiple models, adaptive attacks, property-based testing, concurrency, latency, key compromise, policy change, and independent red-team implementation.

Second, AIS covers codifiable first-line preventive control. Model fitness, monitoring thresholds, compliance judgement, and internal-audit opinion remain separate. Suitability, best execution, market-abuse surveillance, transaction monitoring, and conduct outcomes are often statistical, comparative, or ex-post by construction; they are routed to independent monitoring by design rather than omitted accidentally. The inference provider also remains a critical third party whose concentration, substitutability, exit, and degraded modes require field evaluation.

Third, the FAI and deterministic policy can be wrong, incomplete, stale, or discriminatory. Consistent enforcement does not make a policy legitimate, and a quorum does not prove that an approver understood or was competent. Human exception design can become a bottleneck or bypass. Official cases demonstrate relevance, not AIS adoption or causal improvement. Institutional field studies must examine operating cost, false escalation, recovery, governance, and organizational accountability.

Fourth, an evidence bundle does not automatically establish a complete population. Population testing requires reconciliation between every in-scope execution endpoint and its evidence record, coverage of manual and degraded paths, and detection of omitted or off-system actions. An anchor makes later alteration detectable for included objects; it does not prove that no object was excluded.

Fifth, the Blockscout sample is non-random, based on community-maintained addresses and recent pages, and includes one address-level request failure. Candidate identification is narrower than address activity but does not apply the population-scale event and clustering methods of independent studies. Public addresses cannot reveal legal identity, agent control, commercial relationships, or delivery. The ledger findings therefore characterize independently observable execution evidence and its institutional boundary, not an adoption rate.

\section{Conclusions}

Agentic finance should reproduce the responsibilities of the existing governance architecture while migrating selected human-performed checks to agent-performed or contract-enforced controls. AIS supplies the codifiable first-line preventive mechanism: it separates adaptive inference from the power to make a financial action executable, binds a typed FAI to registered identity, accountable ownership, appetite and mandate lineage, policy, input quality, state, approval, and exact economic semantics, and emits a reproducible authorization reason. Independent validation, monitoring, audit, and legal accountability remain with their institutional owners.

Across 48 fixed prototype fixtures, the tested AIS implementation accepted all admissible cases while preventing all represented authorization, replay, and substitution effects, retaining all 13 defined evidence fields, and detecting every represented service-delivery break. These results demonstrate architecture mechanics within the harness, not production effectiveness. The public-ledger test reaches a dual conclusion: blockchain can independently evidence selected authorization fields, ordering, and settlement, while institutional authority and economic completion require complementary evidence. The encodable--attestable--human taxonomy sharpens the boundary: smart contracts can enforce an approved representation, attestations can carry externally known facts, and people remain responsible for judgement and legitimacy. AIS and blockchain are therefore complementary. AIS decides whether a specific FAI may act; blockchain can make granted authority bounded, executable, and portable across organizations; and institutional governance remains the source of accountability, correction, and legitimacy.

\section*{Abbreviations}
The following abbreviations are used in this manuscript:\\
\noindent
\begin{tabular}{@{}ll}
AIS & Authority--Inference Separation\\
AI & Artificial intelligence\\
EIP & Ethereum Improvement Proposal\\
FAI & Financial action intent\\
FX & Foreign exchange\\
IAM & Identity and access management\\
RBAC & Role-based access control\\
SoD & Segregation of duties\\
USDC & USD Coin
\end{tabular}

\appendix

\section{Prototype Scenario Families and Denominators}

\begin{table}[H]
\caption{Scenario population repeated across four financial domains.\label{tab:fixture-appendix}}
\small
\begin{tabularx}{\textwidth}{p{0.25\textwidth}p{0.16\textwidth}Yp{0.18\textwidth}}
\toprule
\textbf{Scenario} & \textbf{Phase} & \textbf{Expected AIS behavior} & \textbf{Metric population}\\
\midrule
Valid standard / valid with approval & Pre-control & Issue token and execute once; retain complete evidence. & 8 valid fixtures\\
Amount, asset, expiry, stale policy, missing approval, injection & Pre-control & Deny before token issuance with reason. & 24 authorization attacks\\
Recipient or rail substitution & Post-control & Reject exact-intent binding mismatch at adapter. & 8 substitutions\\
Token replay & Replay & Accept seeded first use; reject second nonce use. & 4 replay attempts\\
Service-delivery failure & Post-execution & Preserve execution receipt, withhold completion, record reconciliation break. & 4 delivery failures\\
\bottomrule
\end{tabularx}
\end{table}

\section{Reproducibility and Claim-Boundary Checklist}

\begin{enumerate}[leftmargin=*]
\item Preserve source URL, retrieval time, raw payload, derived transformation, code version, fixture version, and checksum.
\item Report fixture population, exclusions, metric numerators and denominators, and configuration-specific mechanisms.
\item Distinguish mechanism-test outcomes from language-model performance, institutional effectiveness, and real-world attack prevalence.
\item Distinguish facilitator-address activity, candidate authorization calls, verified protocol settlement, authenticated agent initiation, independent counterparties, and delivered service.
\item Do not compare a latest-activity sample's value or concentration directly with population metrics.
\item Link every claimed control outcome to executable fixtures or institutional observations; do not infer effectiveness from architecture diagrams or chain visibility.
\end{enumerate}

\section{Conventional Control Environment Mapped to AIS}

Table~\ref{tab:control-environment-map} maps the conventional control environment to the article's first-line scope and its handoffs. ``First line'' identifies obligations within the AIS design boundary; ``handoff'' identifies information that AIS can produce but independent functions must own.

\footnotesize
\begin{longtable}{p{0.17\textwidth}p{0.13\textwidth}p{0.28\textwidth}p{0.29\textwidth}}
\caption{Conventional control environment mapped to AIS and its residual obligations.\label{tab:control-environment-map}}\\
\toprule
\textbf{Control layer} & \textbf{Position} & \textbf{AIS treatment} & \textbf{Residual obligation or extension}\\
\midrule
\endfirsthead
\multicolumn{4}{c}{\tablename\ \thetable\ continued}\\
\toprule
\textbf{Control layer} & \textbf{Position} & \textbf{AIS treatment} & \textbf{Residual obligation or extension}\\
\midrule
\endhead
\midrule
\multicolumn{4}{r}{Continued on next page}\\
\endfoot
\bottomrule
\endlastfoot
Board and committees & Governing body; handoff & Approval reference, mandate version, new-domain gate, and accountable owner can be bound into the FAI. & Appetite, oversight hierarchy, accountable individual, exception forum, and reporting remain human; material increases may use higher quorum and time-lock.\\
Policies, procedures, and delegation & First line & Mandate registry, policy version, scope, limits, expiry, and signed authority lookup are deterministic inputs. & Policy lifecycle, approval, back-out plan, appetite cascade, and soft tolerance versus hard limit must be governed.\\
Identity and entitlements & First line & Persistent agent ID, declared purpose, owner, one identity per instance, key-custody status, lifecycle, and recertification are recorded. & A key proves credential use, not institutional identity or legal responsibility; provisioning and decommissioning must be enforced.\\
Model risk and evaluation & Second-line handoff & Model/data versions, approved-use field, fixtures, reason codes, and evidence completeness support review. & Independent validation, proposal-quality assessment, performance monitoring, and revalidation on material change remain separate.\\
Context and information quality & First line & Trusted and untrusted sources, model and data versions, alternatives, and an input-quality record enter the FAI. & Retrieval corpus tiering, source-of-record controls, injection boundary, and quality thresholds require operational ownership.\\
Ex-ante transaction control & First line; core & Complete mediation, least privilege, fail closed, canonical intent hash, current policy, expiry, and single-use token. & Retained as the reference standard; the prototype directly tests specified bypass, substitution, and replay effects.\\
Compliance monitoring & Second-line handoff & The evidence bundle exposes the defined in-scope decision population and supports exception, escalation, and break-age analysis. & Monitoring plan, thresholds, population reconciliation, thematic review, and escalation protocol remain independently owned.\\
Internal audit & Third-line handoff & Pinned governance, policy, state, reason, execution, and reconciliation support deterministic re-performance. & Audit coverage, opinion on policy legitimacy, findings, and remediation tracking remain independent human judgements.\\
Four-eyes operations & First line & Human-gate eligibility, checker independence, exact binding, delivery test, and retained reconciliation break. & Eligible approver set, disjoint keys, break ageing, provisioning, escalation, and named write-off authority must be operated.\\
Explanation and contestability & First line plus legal handoff & The control-decision record emits deterministic passed/failed rules and a reason code; model rationale remains non-authoritative. & Human review, complaints, redress, and substantive explanation obligations remain domain- and jurisdiction-dependent.\\
Third-party and resilience & First line & Fail-closed behavior, declared degraded route, manual-action evidence, and reconciliation back to AIS. & Provider concentration, substitutability, exit, impact tolerance, continuity testing, and recovery remain institutional responsibilities.\\
Controls and systems & First line & Reference monitor, scoped token, smart-contract verifier, evidence bundle, hash anchor, revocation and upgrade records. & Key and upgrade governance, revocation propagation, in-flight tokens, retention, erasure, and correction of valid but wrong instructions require explicit policy.\\
\end{longtable}
\normalsize

\section{End-to-End Governance Lifecycle}

Table~\ref{tab:governance-lifecycle} locates the transaction-time AIS core within the full lifecycle from appetite setting to independent assurance and feedback. It distinguishes functions implemented in the artifact from extensions and institutional handoffs.

\footnotesize
\begin{longtable}{p{0.04\textwidth}p{0.09\textwidth}p{0.25\textwidth}p{0.16\textwidth}p{0.30\textwidth}}
\caption{End-to-end governance lifecycle for an agentic financial action.\label{tab:governance-lifecycle}}\\
\toprule
\textbf{Step} & \textbf{Phase} & \textbf{Function} & \textbf{Status} & \textbf{Control or evidence boundary}\\
\midrule
\endfirsthead
\multicolumn{5}{c}{\tablename\ \thetable\ continued}\\
\toprule
\textbf{Step} & \textbf{Phase} & \textbf{Function} & \textbf{Status} & \textbf{Control or evidence boundary}\\
\midrule
\endhead
\midrule
\multicolumn{5}{r}{Continued on next page}\\
\endfoot
\bottomrule
\endlastfoot
1 & Set up & Governing body sets risk appetite and tolerance bands. & Human handoff & The body and accountable owner legitimate the source; the approved reference can be encoded.\\
2 & Set up & Committee approves mandate, scope, ceiling, expiry, and new-domain gate. & Human plus encodable & m-of-n approval and time-lock can enforce the operational representation.\\
3 & Set up & Agent is provisioned, registered, assigned purpose and owner, and given one controlled identity. & v2 design & A credential identifies the actor only when connected to the governed registry and lifecycle.\\
4 & Set up & Policy is approved, released with back-out plan, and its current hash is pinned. & Extended & A verifier can reject superseded policy; approval and policy quality remain institutional.\\
5 & Propose & Context is assembled from tiered sources and input quality is recorded. & v2 design & External content is untrusted by default and retrieval is treated as a control boundary.\\
6 & Propose & Agent emits a typed FAI. & Core artifact & The FAI is the narrow waist between probabilistic inference and deterministic authority.\\
7 & Decide & Independent control plane evaluates governance linkage, mandate, policy, state, and approval. & Core artifact & Returns a token or denial without allowing the inference path to self-authorize.\\
8 & Decide & Deterministic reason code is emitted. & v2 extension & The control record explains authorization; model rationale remains diagnostic.\\
9 & Decide & Eligible and independent human gate is invoked where required. & Extended & Eligibility and segregation are machine-testable; substantive judgement remains human.\\
10 & Decide & Scoped token is issued. & Core artifact & Bound to canonical FAI, agent, mandate, appetite, policy, rail, expiry, and nonce.\\
11 & Execute & Adapter independently re-verifies the token. & Core artifact & Rejects post-check mutation, stale authorization, wrong rail, or invalid credential.\\
12 & Execute & Settlement occurs and nonce is consumed atomically. & Core artifact & Contract constrains the transition and emits receipt and evidence anchor.\\
13 & Execute & Delivery evidence is tested. & Core artifact plus attestation & A signed claim can be verified, but delivery truth may remain external to the chain.\\
14 & Execute & Reconciliation break is aged, escalated, provisioned, and resolved. & v2 extension & A raised but unmanaged break is not treated as a completed control.\\
15 & Evidence & Pre- and post-action bundle is assembled and selectively anchored. & Core artifact & Confidential data remain off-chain; hash and data retention are governed explicitly.\\
16 & Assure & Second line reconciles and tests the eligible decision population. & Institutional handoff & AIS supplies data; monitoring plan, thresholds, challenge, and escalation remain independent.\\
17 & Assure & Third line re-performs decisions and opines on design and legitimacy. & Institutional handoff & Replay proves conformance to pinned inputs, not that the policy was defensible.\\
18 & Feedback & Mandate or policy is revoked or amended and fixtures are rerun after material change. & Extended & Revocation propagation, in-flight tokens, findings, and remediation close the loop.\\
\end{longtable}
\normalsize

\bibliographystyle{plainnat}
\bibliography{references}

\end{document}